\documentclass[12pt]{article}

\usepackage{amsthm}
\usepackage{graphicx}
\usepackage{amsmath}
\usepackage{amssymb}
\usepackage{amsfonts}
\usepackage{appendix}
\usepackage[british]{babel}
\usepackage{slashed}
\usepackage{afterpage}
\usepackage{cite}
\usepackage{color}
\include{epsf}
\usepackage{pdfsync}
\usepackage{color}

\definecolor{verde}{cmyk}{.83,.21,1,.08}

\newcommand{\la}[1]{\label{#1}}

\def\nn{\nonumber}

\newcommand{\Tr}[1]{\:{\rm Tr}\,#1}

\def\be{\begin{equation}}
\def\ee{\end{equation}}
\def\bea{\begin{eqnarray}}
\def\eea{\end{eqnarray}}

\newcommand{\del}{\partial}

\newcommand{\e}{{\mathrm e}}

\newcommand{\eqn}[1]{(\ref{#1})}

\begin{document}

\begin{titlepage}
\begin{flushright}
ICCUB-14-068\\
KCL-PH-TH/2014-51
\end{flushright}

\begin{center}

\baselineskip=24pt

{\Large\bf  Spectral action with zeta function regularization}

\baselineskip=14pt

\vspace{1cm}

{Maxim A.\ Kurkov$^{1,2}$, Fedele Lizzi$^{2,3,4}$, Mairi Sakellariadou$^{5}$, Apimook Watcharangkool$^{5}$
}
\\[6mm]
$^1${\it Dipartimento di Matematica e Applicazioni ``Renato Caccioppoli", \\ Universit\`{a} di Napoli
{\sl Federico II}}
\\[4mm]
$^2${\it INFN, Sezione di Napoli}
\\[4mm]
$^3${\it Dipartimento di Fisica, Universit\`{a} di Napoli
{\sl Federico II}}\\[4mm]
{\it Monte S.~Angelo, Via Cintia, 80126 Napoli, Italy}
\\[4mm]
$^4$ {\it Departament de Estructura i Constituents de la Mat\`eria,
\\Institut de Ci\'encies del Cosmos,
Universitat de Barcelona,\\
Barcelona, Catalonia, Spain}
\\[4mm]
$^5$ {\it Physics Department, King's College London, University of London\\
Strand, London WC2 2RLS, U.K.}\let\thefootnote\relax\footnote{\tiny Max.Kurkov@na.infn.it, Fedele.Lizzi@na.infn.it, Mairi.Sakellariadou@kcl.ac.uk, Apimook.Watcharangkool@kcl.ac.uk}

\end{center}

\vskip 2 cm

\begin{abstract}
In this paper we propose a novel definition of the bosonic spectral action using zeta function regularization, in order to address the issues of renormalizability and spectral dimensions. We compare the zeta spectral action with the usual (cutoff based) spectral action and discuss its origin, predictive power, stressing the importance of the issue of the three dimensionful fundamental constants, namely the cosmological constant, the Higgs vacuum expectation value, and the gravitational constant. We emphasize the fundamental role of the neutrino Majorana mass term for the structure of the bosonic action.
\end{abstract}

\end{titlepage}

\section{Introduction}
The Standard Model  of particle interactions is very successful, and the recent discovery of the Higgs boson seals its validity. Yet there are unanswered questions which on one side suggest to go ``beyond the Standard Model'', perhaps connecting with a theory of quantum gravity, and on the other side beg for an ``explanation'' of the loose conceptual aspects, such as the hierarchy problem or the nature of symmetries. The spectral approach to noncommutative geometry~\cite{connes-book1, connes-book2} provides a framework for the description of the Standard Model encoding it in a general view of geometry based on an algebraic description. 

In particular, the action of a field theory is encoded in such a description, and one can construct a natural action for fermions and bosons based on the spectral properties of the (generalized) Dirac operator. Such a \emph{spectral action} has been introduced in Ref.~\cite{spectralaction}  and applied to the Standard Model in various forms (for a recent review see Ref.~\cite{Walterbook}).  This action is immediately applicable to the phenomenology and has been  presently refined to confront itself with experimental results. Nevertheless, it is not free from conceptual issues. In this paper we will mostly deal with the latter and try to solve some of these drawbacks with the introduction of a new form of action, the \emph{$\zeta$ spectral action}. In the following, in order to distinguish the $zeta$ spectral action from the usual one, we will call the latter the \emph{cutoff spectral action} since the main difference lies in the regularization procedure. 

The key difference between the two actions lies in the fact that no operators of dimension higher than four appear in the $\zeta$ spectral action, and therefore the theory is \emph{renormalizable}. In particular, it is not necessary to have to consider it as an effective theory valid just below the unification scale, and one can safely use it up to the Planck scale where  the very nature of spacetime changes due to quantum gravitational effects.
The ultraviolet asymptotics of the cutoff spectral action was discussed in Ref.~\cite{Kurkov:2013kfa}, finding the non propagation of  bosons.  In Ref.~\cite{Alkofer:2014raa} it was shown  that all spectral dimensions coming from the cutoff based bosonic spectral action do not coincide with the topological $d = 4$, viz. all of them are zero, implying that some sort of ultraviolet completion, like asymptotic safety~\cite{Niedermaier:2006wt}, is necessary. The $\zeta$ spectral action instead exhibits viable spectral dimensions.  For Higgs scalars and gauge bosons the spectral dimensions coincide with the topological one and equal four, while in the gravitational sector the spectral dimension equals two, which implies improved ultraviolet behavior of the gravitational propagators.

The $\zeta$ spectral action defines naturally, at the classical level, all dimension four operators, therefore the crucial point is the generation of  lower dimensional  operators, namely the quadratic Higgs term, and the Einstein-Hilbert term if one considers the Standard Model coupled to gravity. It turns out, and this is one of the most interesting aspects of the paper, that these terms are generated by the presence, in the Dirac operator, of a term corresponding to the right handed neutrino Majorana mass. It is already known~\cite{Stephan, Resilience, coldplay, Chamseddine:2013rta, Corfu,twistedtriple} that the presence of a term (or a field) of this kind is crucial in order to obtain the experimental  Higgs mass. For the current study, this term must contain dimensionful constants, and this leads to the introduction of the needed lower dimensional operators in the spectral action. Compared with the cutoff based approach the zeta spectral action has at the current stage the same predictive power, however the perspective on the hierarchy of scales and the naturalness is conceptually different.  

This paper is organized as follows: In Sec.~\ref{SA} we review the traditional spectral  approach and list its successes and open issues. In Sec.~\ref{ZSA} we introduce an altrenative definition based upon the $\zeta$ function in an attempt to address and cure some of the drawbacks mentioned in the previous section and  discuss in detail the issue of lower dimensional operators and normalization, emphasizing the advantageous aspects of the $\zeta$ spectral action.  Sec.~\ref{GSD} is devoted to the spectral dimensions, and Sec.~\ref{last} contains conclusions and outlook.

\section{The Cutoff Bosonic Spectral Action}\label{SA}
\setcounter{equation}{0}
In noncommutative geometry and its applications for particle physics the basic object is the spectral triple $\mathcal{(A,H,}D)$.
The topology of spaces is described by a (possibily non commutative) $*$-algebra $\mathcal A$, represented as operators on a Hilbert space $\mathcal H$ of spinors, considered as the matter fields of the theory. The geometry is encoded in the (generalized) Dirac operator $\mathcal D$ which contains indiffation on the differential structure of spacetime, and its metric~\cite{connes-book1, connes-book2}. The Standard Model (SM) of strong and electroweak interactions can be explained from a purely geometric approach, considering an ``almost commutative geometry'', namely the product of an ordinary manifold $M$  (representing spacetime) times an internal space described by a matrix algebra, with a particular choice of the algebra $\mathcal A$. To obtain the SM the minimal choice of $\mathcal A$ is the algebra product of smooth functions on $M$  times the finite algebra of complex numbers plus quaternions plus three by three matrices~\cite{Chamseddine:2006ep}:
\be
\mathcal A= C^{\infty}(M)\otimes (\mathbb C\oplus \mathbb H\oplus M_3(\mathbb C))~.
\ee
The Dirac operator comprises a continuous part acting on functions of spacetime, times a finite dimensional part which contains the information of the masses and mixings of the physical fermions:
\be
D_0=(\slashed \del+\slashed\omega) \otimes \mathbb I + \gamma_5\otimes D_F~,
\ee
where $\omega$ is the spin connection, $\gamma_5$ is the usual product of the four $\gamma$ matrices and $D_F$ is a matrix containing the masses (or rather Yukawa coulings) of the fermions. The covariant version of the Dirac operator is built with the addition of a generic  connection\footnote{More precisely it should be $D=D_0+A+JAJ$, but the real structure $J$, which is otherwise very important for physics, plays no role in this study, hence we ignore it to simplify notation.} 
\be 
D=D_0+A~,
\ee
where  $A=\sum_i a_i [D,b_i]$, with $a_i,b_i$ elements of the algebra, is the algebraic representation of the potential connection one-form. 

In the cutoff approach the bosonic spectral action reads\footnote{The subscript $\Lambda$ to the spectral action is needed to differentiate it from the $\zeta$ spectral action which we introduce below, and indicate as $S_\zeta$.}
\bea
S_{\Lambda}= \Tr \left(\chi{\left(\frac{D^2}{\Lambda^2}\right)}\right)~, \label{bsadef}
\eea 
with $\chi$ a cutoff function, and the scale $\Lambda$ is some constant. This cutoff function is one at the zero value of its argument, and vanishes at $\infty$. 
Typical examples considered in the literature are $\chi(x)=1$ if $x\leq\Lambda$ and zero otherwise, or a smoothened version of it, alternatively one can consider an exponentially decreasing cutoff $\chi(x)=\e^{-x}$.  

Both $\chi$ and $\Lambda$ are needed to define a finite trace using  eigenvalues of the Dirac operator, but they have to be considered as inputs. In particular $\chi$ enters in the heat kernel expanded action via its momenta, which are undefined in the theory\footnote{The momentum of the cutoff function associated with the gauge couplings at unification has been constrained by astrophysical data~\cite{Nelson:2010rt,Nelson:2010ru, Lambiase:2013dai, Capozziello:2014mea}.}. Our work starts from the observation that the introduction of a cutoff function and a scale is not the only way in order to construct a finite trace using the Dirac operator. This turns out to be a crucial starting point for the different version of the spectral action, the $\zeta$ action we will present in the next section.

The spectral action is a classical quantity and can be calculated using the heat kernel expansion. The result is a polynomial in decreasing powers of the cutoff scale $\Lambda$ starting from $\Lambda^4$. The resulting terms depend on the momenta of the cutoff function $\chi$ and the fields which appear in the corresponding powers of $D$, and their derivatives.
The obtained result can schematically be written as~\cite{Chamseddine:2006ep}
\bea
S_{\Lambda} &=& \int d^4x \sqrt{g} \left( A_1 \Lambda^4  \nonumber\right.\\&& + A_2 \Lambda^2\left(\frac 54 R - 2 y_{\rm t}^2H^2 - M^2 \right)  \nonumber\\&& +A_3\left( g_2^2W^{\alpha}_{\mu\nu}W^{\alpha~\mu\nu} + g_3^2 G^a_{\mu\nu}G^{a~\mu\nu} + \frac{5}{3}g_1^2 B_{\mu\nu}B^{\mu\nu}\right)\nonumber\\&& + \mbox{other ${\cal O}(\Lambda^0)$ } \nonumber\\&&\left. +  {\cal O}\left(\Lambda^{-2}\right) \right)~, \label{sa}
\eea
where $y_{\rm t}$ is the Yukawa coupling for the top quark\footnote{The action contains terms depending on all fermion's Yukawa couplings, which however can be ignored given the top predominance.}; $R$ is the curvature scalar; and $W,G,B$ are the curvature tensors for the three interactions and $M$ is (up to a numerical factor)  a heavy Majorana right handed neutrino mass (which would appear also in the $\Lambda^0$ term variously coupled).  
The $A_i$ with $i=1,2,3$ are constants which depend on the details of the function $\chi$ and for typical choices of that function the three constants are not too different from unity. In the case of the cutoff being the characteristic function of the unit interval the ${\cal O}(\Lambda^{-2})$ are not present in the asymptotic expansion, however careful analysis of the situation shows~\cite{Iochum:2011yq} that the anzatz
\eqref{sa} coincides with the left hand side only for momenta smaller than the cutoff $\Lambda$. 
 
In Eq.~\eqref{sa} the  $g_1$, $g_2$, $g_3$ are the corresponding gauge couplings of the three interactions. The action we have written is classical, but the relation among the $g_i$'s indicates that the action $S_\Lambda$ is written at the scale at which  all three gauge constants are equal (up to the usual factor of $5/3$ normalizing the abelian interaction): 
\bea
\frac 53 g_1^2(\Lambda) = g_2^2(\Lambda) = g_3^2(\Lambda)~. \label{unif}
\eea
Present data indicate that this is not true, at least in the absence of new physics, but experimentally it results that the three couplings are very close to each other in the energy range of $(10^{14}-10^{17})$~GeV. Modification of the action, as for instance within a supersymmetric theory~\cite{susy}, or by considering the six dimensional terms of order $\Lambda^{-2}$~\cite{Nomadi}, may lead to a more concrete unification.

The presence of a unification point does not necessarily imply a larger, grand unified group. In fact the most common scenarios, based on SU(5) or SO(10) do not fit in the noncommutative geometry frame, although there is strong indication that some sort of Pati-Salam symmetry may be present~\cite{Chamseddine:2013rta}. Nevertheless, independently of the choice of gauge group among \emph{allowed}, the cutoff spectral action requires the presence of an additional scale $\Lambda \sim(10^{14}-10^{17})$~GeV.

The presence of the unification point of the three interactions is fundamental to the theory, and does not depend on the fact that a regularization is needed, in other words it does not depend on the form of $\chi$ or the value of  $\Lambda$. Its specific value is related to the spectral data contained in $D$, and one might say that its presence is more of kinematical nature. This observation is important for the alternative definition that we advocate, which exploits the spectral triple data only.  
We will first however discuss the successes and open issues of the cutoff bosonic spectral action.

\subsection{Why the Spectral Action}

From the symmetry point of view the main point  is that very few gauge groups fit into the model. Fortunately the standard model one does, but very few others do. For example the Pati-Salam group SU(2)$_L\times$ SU(2)$_R\times$ SU(4) does, but SO(10), of which the former group is an intermediate breaking stage, does not. Moreover the formalism allows fermions in the fundamental or trivial representation of the gauge group, a feature of the Standard Model. The absence of larger groups with novel representation prevents proton decay.

The main success of the spectral action is however the possibility to infer quantities related to the boson (and the Higgs in particular) based on the input of only fermionic parameters (Yukawa couplings and mixing) present in the (generalized) Dirac operator. The Dirac operator defines also the fermionic part of the action as a structure of bilinear form acting on fermions:
\be
S_{\rm F} = \langle J \psi |D |\psi \rangle~.
\ee

As we said the model has predictive power, and the initial prediction for the mass of the Higgs was 170~GeV~\cite{Chamseddine:2006ep}; it is now known this number is not correct. It is nevertheless remarkable that a theory based on mathematical first principles quantitatively predicts a number which is not too far from the experimental one. Taking the input from experiment the models can be improved with the introduction of a scalar field which alters the running of the quartic coupling and makes it compatible with the actual mass of 126~GeV~\cite{Resilience, coldplay, Chamseddine:2013sia, Chamseddine:2013rta}.
This scalar, usually called 
$\sigma$ in this context,  had appeared before within noncommutative geometry~\cite{Stephan} as well as in general, see e.g.\ Ref.~\cite{sigmafield}. Although in Ref.~\cite{Resilience}
    it was inserted ad-hoc in the spectral action
	principle, there are approaches~\cite{coldplay}, where it comes naturally via the formalism based on introduction of the so called grand symmetry. Alternatively, it is possible a violation of some of the non commutativity conditions, as done in Refs.~\cite{Chamseddine:2013sia, Chamseddine:2013rta}.

\subsection{Why go beyond}
The bosonic spectral action defined by Eq.~\eqref{bsadef} still leaves open some important issues, which we now discuss. 

\begin{figure}[htb]
\begin{center}\includegraphics[scale = 0.45]{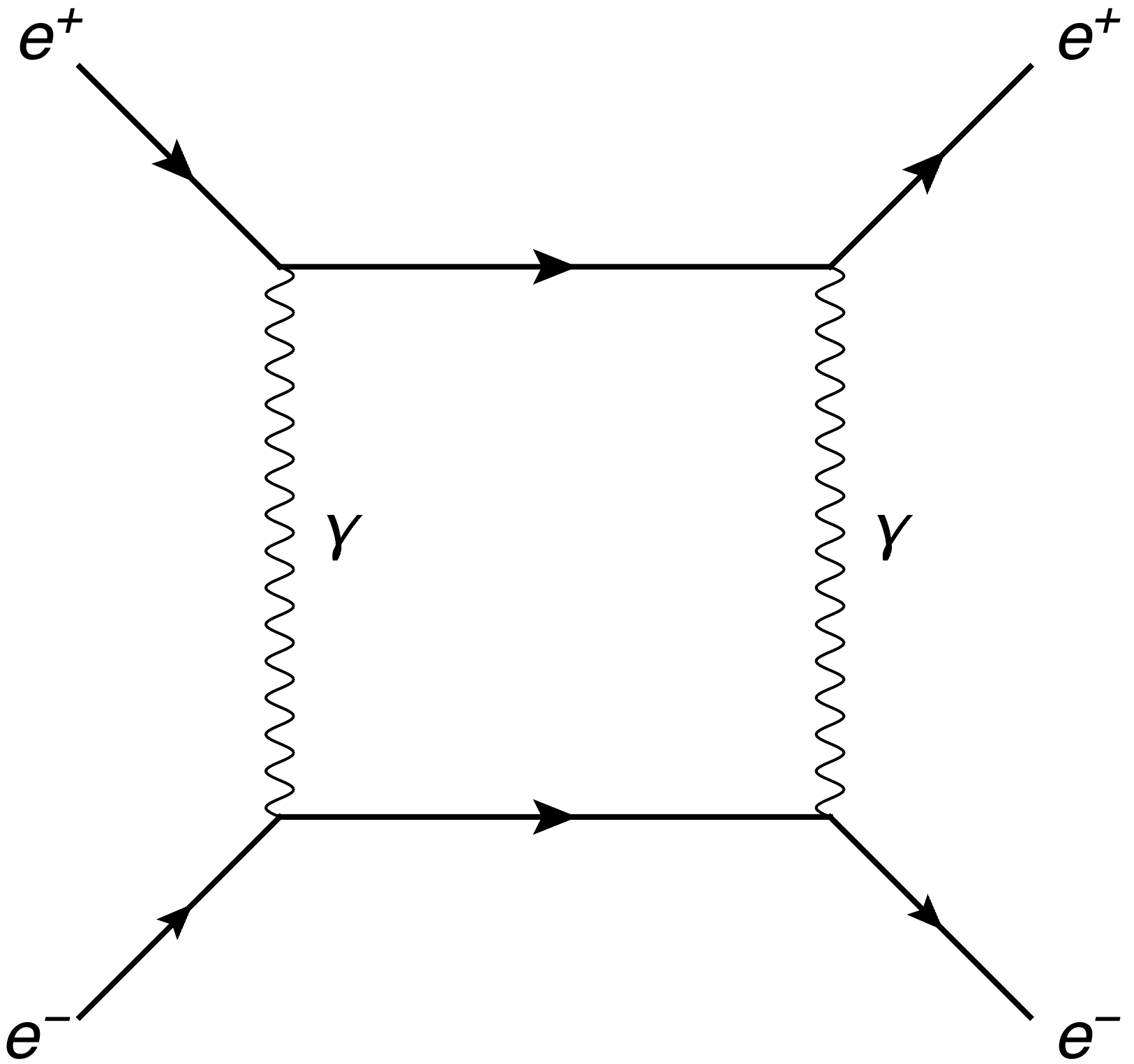}\end{center}
\caption{\sl We present an ultraviolet divergent diagram leading to the introduction in the theory of  four fermionic vertex, i.e. making it nonrenormalizable. Wavy lines present
photon propagators, arrowed lines correspond to electrons and positrons.
}\label{bosonfig}
\end{figure}

The locality of the theory is far from settled. Although in the low momentum regime the expansion~\eqn{sa} recovers the Standard Model action, the high momentum regime does not contain positive powers of the field derivatives~\cite{Iochum:2011yq, Kurkov:2013kfa},
exhibiting the structure 
\be
S_\Lambda\sim \int d^4 x \left(\alpha_1\, \Lambda^2 \,h_{\mu\nu} \, h^{\mu\nu} +\alpha_2 \,\phi \,\frac{\Lambda^4}{\partial^2}\,\phi  
+ \alpha_{3}\,A_{\mu}\,\frac{\Lambda^4}{\partial^2}\,A^{\mu}\right)~,
\ee
where $\phi$ and $A_{\mu}$  are bosons of spin $0$ and $1$ respectively;  $\alpha_{1,2,3}$ are constants
depending on the particular realization of the model. The traverse and traceless fluctuations $h_{\mu\nu}$ of the metric tensor $g_{\mu\nu}$ are defined as follows
\be
g_{\mu\nu} = \delta_{\mu\nu} + \frac{h_{\mu\nu}}{M_{\rm Pl}},  \la{gfluct}
\ee
where $M_{\rm Pl}$ is the Planck mass, i.e.\ they have canonical dimension of energy.
 This opens the question of the meaning of the scale $\Lambda$, and what happens beyond it. Whether the theory is unitary or causal, for example. The theory is certainly not renormalizable, as it stands, since at high momenta the bosonic propagators do not decrease. For instance, in contrast to conventional QED, the diagram presented in Fig.~\ref{bosonfig} is divergent, 
therefore one has to add four fermonic interaction $\left(\bar\psi\psi\right)^2$ in order to subtract the infinity. Theories with four fermonic interactions are well known to be nonrenormalizable. 

The spectral action \eqref{bsadef} is calculated  via the asymptotic heat kernel
expansion. This  can be divergent and generally speaking does not coincide with the spectral action~\cite{Iochum:2012bu}. 

Let us note that doing a rigorous analysis of the heat kernel expansion one does not actually obtain the Higgs potential from the heat expansion, i.e.\ one does not find a minimum for all natural\footnote{Under natural choices of the cutoff function,  we mean a non increasing  cutoff function $\chi(z)$ that equals to 1 at $z=1$ and then
rapidly decreases at $z>1$.} choices of the cutoff function. Indeed, although the finite ansatz based on the first
three nontrivial heat kernel coefficients reproduces familiar double well potential, the total sum does not.
Already for the exponential cutoff the Higgs potential defined as follows
\be
V(H) = \lim_{{\rm vol \rightarrow \infty}} \frac{{ S_{\Lambda}}(H)}{\rm{vol}}, \quad H = {\rm const.}, 
\quad g_{\mu\nu} = \delta_{\mu\nu}, \quad B_{\mu\nu} = G_{\mu\nu} = W_{\mu\nu} = 0~,
\ee
can be computed exactly using results of \cite{Avramidi:1995ik},
and the answer reads
\be
V(H) = \sum_{j=1}^{N_F} B_j \exp{\left(- \frac{\beta_j H^2 }{\Lambda^2}\right)}, \la{Vexact}
\ee
where $B_j$ and $\beta_j$ are nonnegative  constants and $N_F$ is number of fermmions. For a non exponential cutoff, we find that performing  a Laplace transform 
of the cutoff function, the result remains qualitatively valid for all natural cutoff functions. The result \eqref{Vexact} is valid for all $H$, in particular in
``the low energy" region $H \ll \Lambda$ where the cutoff spectral action is supposed to work. Therefore, in order to make this model viable it is necessary to add, by hand,  to the $H^2$ term, a quadratic term with a large coefficient.
Such a term will provide the minimum of the overall potential, while it will lead to the correct Higgs vacuum expectation value, which is many order of magnitude smaller than the cutoff scale $\Lambda$.
Regarding the cosmological constant, let us note that the natural value obtained through the spectral action is $\sim \Lambda^4$, hence much bigger than the observationally known. Hence, one should add by hand the appropriate term to render it compatible with its observational value. Finally it is known,
that the coefficient in front of the scalar curvature $R$, that has to be identified with inverse gravitational constant, is (depending on $\Lambda$) at least
one order of magnitude smaller than the value, obtained from experiment, therefore one should also add a corresponding term to the action 
in order to get the correct value of this coefficient. 
Thus, we conclude that the magnitude of the \emph{dimensionful} parameters appearing in the model, the cosmological constant, the Higgs vacuum expectation value and the gravitational coupling have to be put in~\eqref{bsadef} by hand with unnatural numerical values which are independent of the cutoff scale. We emphasize that independently of the choice of the almost-commutative manifold, the physical values of these quantities necessitate an experimental input which goes beyond the data encoded by the spectral triple.
All these quantities have to be substituted by a subtraction point which fixes their value by hand to fit the experimental data. This drawback is closely related to the naturalness problem.

There are other issues, like the signature\footnote{Alternative approaches based on the
Lorenzian signature were discussed in Refs.~\cite{Franco:2012er,Franco:2013gxa,Besnard:2014rma}.}, or the compactness of spacetime, which are beyond the scope of the present study.

\section{A new proposal: the $\zeta$ spectral action}
\label{ZSA}

Going back to the origins of the bosonic spectral action, one notes that this is a regularized version of the number of eigenvalues of the square of the Dirac operator. The number of eigenvalues of an unbounded operator is of course infinite and one has to (classically) regularize this sum, which would be otherwise $1+1+1\ldots$. The spectral action does it with the introduction of the cutoff scale $\Lambda$.

Since the rationale for introducing the spectral action was a regularization, we may try another regularization of the number of the eigenvalues of an operator. To cure some of the drawbacks of the conventional bosonic spectral action outlined above, we propose a definition of the bosonic spectral action based upon the $\zeta$ function.
We hence define the bosonic spectral action as
\be 
S_\zeta \equiv \lim_{s\rightarrow 0} \Tr D^{-2s}\equiv \zeta(0,D^2)~. \la{szetadef}
\ee
The $\zeta$ function is well defined~\cite{Minakshisundaram:1949xg} and given by the $a_4$ 
heat kernel coefficient associated with the Laplace type operator $D^2$, namely
\bea
S_\zeta &=& a_4\left[D^2\right] = \int d^4 x \,\sqrt{g}\,L~,\nonumber\\
\mbox{with} \quad L(x) &=& a_4(D^2,x)~. \label{BSAv20}
\eea
We refer the reader to e.g., Ref.~\cite{Vassilevich:2003xt} for details. Strictly speaking the trace of $D^{-2s}$ is convergent if the real part of $2s$ is greater than the dimension. Then one can prove that it has a unique meromorphic extension, denoted $\zeta \left(s,D^2\right)$, which has a set of poles on the complex plane of the variable $s$. 
For a Laplace type operator $D^2$, the point $s=0$ is not a pole, therefore $\zeta \left(0,D^2\right)$ is well-defined. The zeta function and its regularity at zero were discussed in \cite{Connes:2006qj} 
in the context of almost commutative manifolds, and in \cite{Eckstein} in the more general noncommutative setup. 
Here we use the  zeta function to regularize the sum in Eq.~\eqref{szetadef} defining the classical action, while in a slightly different context the zeta function regularization is also commonly used  to regularize functional determinants appearing upon quantization~\cite{Hawking}. 
The spectral action~\eqref{BSAv20} is nothing but the conformal anomaly in a theory of quantized fermions~\cite{Vassilevich:2003xt} where the bosonic fields are a classical background, the relation between the cutoff spectral action and the anomaly can be found in Refs.~\cite{AndrianovLizzi, AndrianovKurkovLizzi, KurkovLizzi, KurkovSakellariadou}. 
The Lagrangian density obtained from the $\zeta$ spectral action has the form:
\bea
 L(x) &=& \alpha_1M^4 + \alpha_2 M^2 R +\alpha_3M^2H^2  \nn\\
&&+ \alpha_4 B_{\mu\nu}B^{\mu\nu} + \alpha_5 W_{\mu\nu}^{\alpha}W^{\mu\nu\,\alpha} + \alpha_6 G_{\mu\nu}^aG^{\mu\nu\,a} \nn\\
&& + \alpha_7\,H\left(-\nabla^2-\frac{R}{6}\right)H + \alpha_8 H^4 + \alpha_9 C_{\mu\nu\rho\sigma}C^{\mu\nu\rho\sigma}
+ \alpha_{10}R^*R^*~, \label{L}
\eea
where $B_{\mu\nu}$, $W_{\mu\nu}$ and $G_{\mu\nu}$  are the field strength tensors of the corresponding  U(1), SU(2) and SU(3) gauge fields; $\alpha_{1},..,\alpha_{10}$ are dimensionless  constants determined by the Dirac operator (whose particular form is not relevant here); $R^*R^*$ is the Gauss-Bonnet density and $C$ is the Weyl tensor.

The bosonic spectral action $S_\zeta$  contains only terms needed for the Standard Model and Einstein gravity and \emph{nothing else} (e.g.\ higher dimensional operators) therefore it
is \emph{local}, \emph{renormalizable} and  \emph{unitary}.  This means that one can use renormalization and safely compute an arbitrary loop order corrections. In this analysis, as in general for the spectral action, gravity is a background field which is \emph{not quantized}, therefore there are no issues of renormazability or unitarity as far this sector is concerned.  Another strong advantage of the definition \eqref{BSAv20}
is the fact that the Lagrangian \eqref{L} is an exact result, therefore there is no need to consider asymptotic expansions and their convergence. 

Substituting the Weyl square and Gauss-Bonnet density expressions via  $R^2$, $R_{\mu\nu}R^{\mu\nu}$ and $R_{\mu\nu\alpha\beta}R^{\mu\nu\alpha\beta}$
we can rewrite our Lagrangian as a linear combination 
\bea
&& L(x) = \sum_{j=1}^{12} \eta_{j}O_{j},
\eea 
where 
\bea
&& O_1 = 1, \quad O_2 = R,\quad O_3 = H^2, \quad O_4 = B_{\mu\nu}B^{\mu\nu},\quad O_5 = W_{\mu\nu}^{\alpha}W^{\mu\nu\,\alpha}, \nn \\
&& O_6 = G_{\mu\nu}^a G^{\mu\nu\, a}, \quad O_7 = H\nabla^2H,\quad O_8 = H^2 R, \quad O_9 = H^4,\quad O_{10} = R_{\mu\nu\alpha\beta}R^{\mu\nu\alpha\beta}, \nn\\
&& O_{11} = R_{\mu\nu}R^{\nu},\quad O_{12} = R^2~. \la{genth}
\eea
	The Lagrangian given by Eq.\eqref{genth} is the most general \emph{renormalizible} Lagrangian for QFT in curved spacetime\footnote{For renormalization of QFT in curved spacetime and corresponding counter terms see e.g.\ Ref.~\cite{Brown:1980qq}.},
and correspondingly the complete spectral action 
\be
S = \langle \bar\psi |D|\psi\rangle + S_\zeta~,
\ee
is a renormalizable theory describing the Standard Model. Upon quantization all twelve composite operators $O_j$ in Eq.~\eqref{genth} must be renormalized and after proper introduction of the renormalization matrix and counter terms  the coefficients $\eta_j$  by the end of the day must be replaced by renormalized physical parameters $\eta_j^{\rm phys}$.
Quantum field theory never predicts the physical values of the coefficients $\eta^{\rm phys}_j$ and they must be fixed at some energy scale by normalization conditions.  Usually such normalization is done using the values obtained from experiment at low energy\footnote{Here ``low" may mean TeV scale, which is still much lower than the unification scale.}. For the spectral action however it is natural to fix the scale at the unification point, and this fixes the relations with all other coefficients, which likewise are normalized at the unification point, \emph{with their value given by the spectral action}.  We emphasize that this normalization procedure is not a consequence of the spectral geometry framework, but is a natural prescription. This prescription
gives predictive power, and we will use it considering the scale at which the $\zeta$ spectral action is written to be $\sim (10^{14}-10^{17})$ GeV.
In analogy with the conventional bosonic spectral action discussed in Sec.~\ref{SA}, we will still call this scale $\Lambda$. In conclusion, the bosonic spectral action is written as an action valid at a particular scale, whilst the action is itself independent of this scale.

The spectral approach is very successful giving restrictions on \emph{dimensionless} parameters like Higgs quartic coupling, gauge couplings and Yukawa couplings, etc. For the $zeta$ spectral action in its present formulation the issue is the value of the dimensionful constants in the lower dimensional terms in the action. We notice from~\eqref{L}  that the presence of the Majorana mass term in the Dirac operator introduces the correct lower
dimensional operators, however the corresponding coefficients are physically inappropriate. Therefore these three numbers can
not be taken from the spectral action, and one has to normalize the lower dimensional operators by hand, thereby leaving the
naturalness problem unsolved.

Let us comment here on the fundamental role played by the dimensionful constant $M$ appearing in the position corresponding to the Majorana mass in the Dirac operator. At this stage this is a constant quantity, at the end of the paper we comment on the interplay between this term and the $\sigma$ field. 
As we said, bare values of the dimensionful parameters must be renormalized by hand.  However, the corresponding terms in~\eqref{L} carry information: they define the structure of the counter terms needed to eliminate
divergences upon quantization when one uses dimensional regularization. 
		Indeed, if one has $M = 0$,  since there are no dimensionful constants in the bare Lagrangian anymore, divergences proportional to $1$, $R$ and $H^2$ would not appear, and  there would be no necessity to introduce the corresponding counter terms. Correspondingly, the cosmological constant, Higgs mass parameter and the gravitational constant would never come out from renormalization. 
In the context of the spectral action the Majorana mass term already plays a fundamental role for the phenomenological viability of the model; in the present context its role is even enhanced.

\noindent{{\bf Remark} \sl \small It is useful to compare our approach with the one of Ref.~\cite{vanSuijlekom:2011kc} where the spectral action was defined by the ansatz of the (generally divergent)
asymptotic expansion
\be
\sum_{n=0}^{N} f_{2n}  \Lambda^{4-2n}a_{2n}[D], \la{aesa}
\ee 
where $f_n$ are arbitrary and $N\geq2$. This makes the theory local and super-renormalizable, with $\Lambda$ a cutoff, not a physical scale.

The higher terms are a particular kind of higher derivative regularization \cite{Slavnov:1971aw}, in particular when $N=3$ we have the following action for the gauge field
\be
f_{4}F_{\mu\nu}^a F^{\mu\nu~a} + \frac{f_6}{\Lambda^2} F_{\mu\nu}^a (-\partial)^2 F^{\mu\nu~a},
\ee
which improves the ultraviolet behaviour of the propagator
\be
\frac{1}{p^2} \rightarrow \frac{1}{p^2 + \frac{f_6}{f_4\Lambda^2}p^4}.
\ee
At finite values of $\Lambda$ such theories are known to be super renormalizable (but with ghosts) and in the limit $\Lambda\rightarrow\infty$ one recovers the original renormalizable (without ghosts) theory. Since there are still divergent one loop fermionic diagrams one would then have to regularize the theory with dimensional regularization, thereby creating an artificial hybrid of higher-derivative  and dimensional regularizations \cite{Martin:1994na}.
For $N = 2$ in flat spacetime the action is renormalizable and unitary. However, the coefficients $a_0$ and $a_2$ that are supposed to introduce the cosmological constant, Higgs vacuum expectation value and Einstein-Hilbert action term  do not have by themselves predictive power, since all
these parameters have to be normalized using experimental values.  If, keeping $N=2$, one removes them by hand, the definition \eqref{aesa} will lead to our definition~\eqref{szetadef}.$\blacksquare$}

\section{Gravitational Spectral Dimension}
\label{GSD}
\setcounter{equation}{0}
Unlike the cutoff formulation our new definition leads to nontrivial spectral dimensions,
which we calculate in this section. The spectral dimension is conventionally the effective dimension of the manifold probed by the particles ``living" on it (see for instance Ref.~\cite{Alkofer:2014raa} and references therein for details). In particular, for the standard action of a particle with mass $m$, one has to replace $p^2 - a p^4$ with $p^2 +m^2$ in
\eqref{hk}, leading to a spectral dimension which will coincide with the topological one, namely $d=4$. For a more complicated momenta dependence, $D_{\rm s}$ can be different from the topological dimension, being dependent on the particles one chooses to probe, since their propagators can have different dependence on the momenta.

Since the actions for Higgs scalar and for gauge fields have the same behavior in the ultraviolet, like in the infrared, their corresponding
spectral dimensions coincide with the topological dimension of the manifold and are equal to four. The gravitational spectral dimension can be 
also defined in a viable way, however such a definition requires some analytical continuation, therefore we elaborate carefully this point.

The gravitational part of our theory consists of the Weyl square contribution coming from Eq.~\eqref{BSAv20} and the Ricci scalar $R$ appearing after the renormalization discussed in the previous section:
\bea
S_{\rm gr} = \int d^4 x \sqrt{g} \left( \frac{M_{\rm Pl}^2}{16 \pi}R  - \frac{N_F}{16\pi^2} C_{\mu\nu\eta\xi}C^{\mu\nu\eta\xi}\right), \label{Sg}
\eea
where $N_{\rm F}$ stands for the number of fermions.

To compute the spectral dimension one has to extract the quadratic part of $S_{\rm gr}$ for transverse and traceless fluctuations $h_{\mu\nu}$ of the metric tensor $g_{\mu\nu}$ defined by \eqref{gfluct}.
We obtain
\bea 
S_{\rm gr} = \frac{M_{\rm Pl}^4}{64\pi} \int d^4x \,h_{\mu\nu}\left[\left(-\partial^2\right) - a\left(-\partial^2\right)^2\right]h^{\mu\nu}
+{\cal O}(h^4) \label{Sg2}~,
\eea
where  
\bea
a\equiv \frac{2N_F}{\pi M_{\rm Pl}^2}~.
\eea
To define the spectral dimension one needs the heat kernel $P(T,x,x')$ corresponding to 
Eq.~\eqref{Sg2},
 or more precisely
its value $P(T)$ at $x = x'$.
One can see from Eq.~\eqref{Sg2} that such a heat kernel is given by
\bea
P(T,x,x') = \int \frac{d^4 p}{(2\pi^4)}e^{ip(x-x')} e^{-\left(p^2 - a p^4\right)T}~.\label{hk}
\eea
Note that setting $x= x'$, the integral $P(T)$ is divergent, because $a$ is positive
whilst is well defined for negative $a$. 
In spherical coordinates the relevant integral is proportional to
\bea
 \int_0^{\infty} dp\, p^3 e^{-\left(p^2 - a p^4\right)T}
=\frac{1}{8}\,\frac{ \left( 2\sqrt{-a\,T}\,{{\exp}{\left({\frac {T}{4a}}\right)}} -
\sqrt {\pi }\,
{{\rm erf}\left(\frac{1}{2}{{\frac{\sqrt{-a\,T}}{a}}}\right)}T-\sqrt {
\pi }T \right) {{\rm e}^{-{\frac {T}{4a}}}} }{\left( -a \,T
 \right) ^{\frac{3}{2}}}~, 
\eea
where the right hand side is an analytic function on the complex plane without a ray, that we can choose as a lower half of imaginary axis $[0,-i\infty)$.
It means that there exists an \emph{analytic continuation} in a region of positive $a$; we \emph{define} our
integral for positive $a$ as such an analytic continuation.

We are now ready to compute the spectral dimension
\be
D_{\rm s} \equiv \lim_{T\rightarrow 0}\left[-2 \frac{\partial \log P(T)}{\partial \log T}\right]~. \label{spdim}
\ee
Following Ref.~\cite{Alkofer:2014raa}, apart from the spectral dimension we also introduce the ``running" spectral dimension $\widetilde {D_{\rm s}}(T)$
\be
\widetilde {D_{\rm s}}(T) \equiv  -2 \frac{\partial \log P(T)}{\partial \log T}~, \label{runsd}
\ee
which in our case is imaginary, however its limit (i.e.\ the \emph{conventional} spectral dimension) we will show that is real. 
We write
\be
\widetilde {D_{\rm s}}(T)= \frac{ 2\sqrt{-a\,T} (4 a+T){\exp{\left({\frac {T}{4a} }\right)}} -T(2a+T)
\sqrt {\pi}
\left[{{\rm erf}\left({{\frac{1}{2}\, \frac{\sqrt{-a\,T}}{a}  }}\right)} + 1\right]
}
{2 a \left(2\sqrt {-a\,T}\,{\exp{\left({\frac {T}{4a} }\right)}}
-\sqrt {
\pi }{\rm erf}\left({{\frac{1}{2}\, \frac{\sqrt{-a\,T}}{a}  }}\right)T
-
\sqrt {\pi }\,T \right) 
}~,
\ee
and plot  $\widetilde {D_{\rm s}}(T)$ in Fig.~1. Although we are interested in the limit 
$T\rightarrow 0$, it is worth to note that in the limit $T\rightarrow +\infty$ the ``running" spectral
dimension is real, while it would be interesting to understand the meaning of the ``running"
imaginary spectral dimension. 

\begin{figure}[top]\begin{center}
\includegraphics[scale = 0.4]{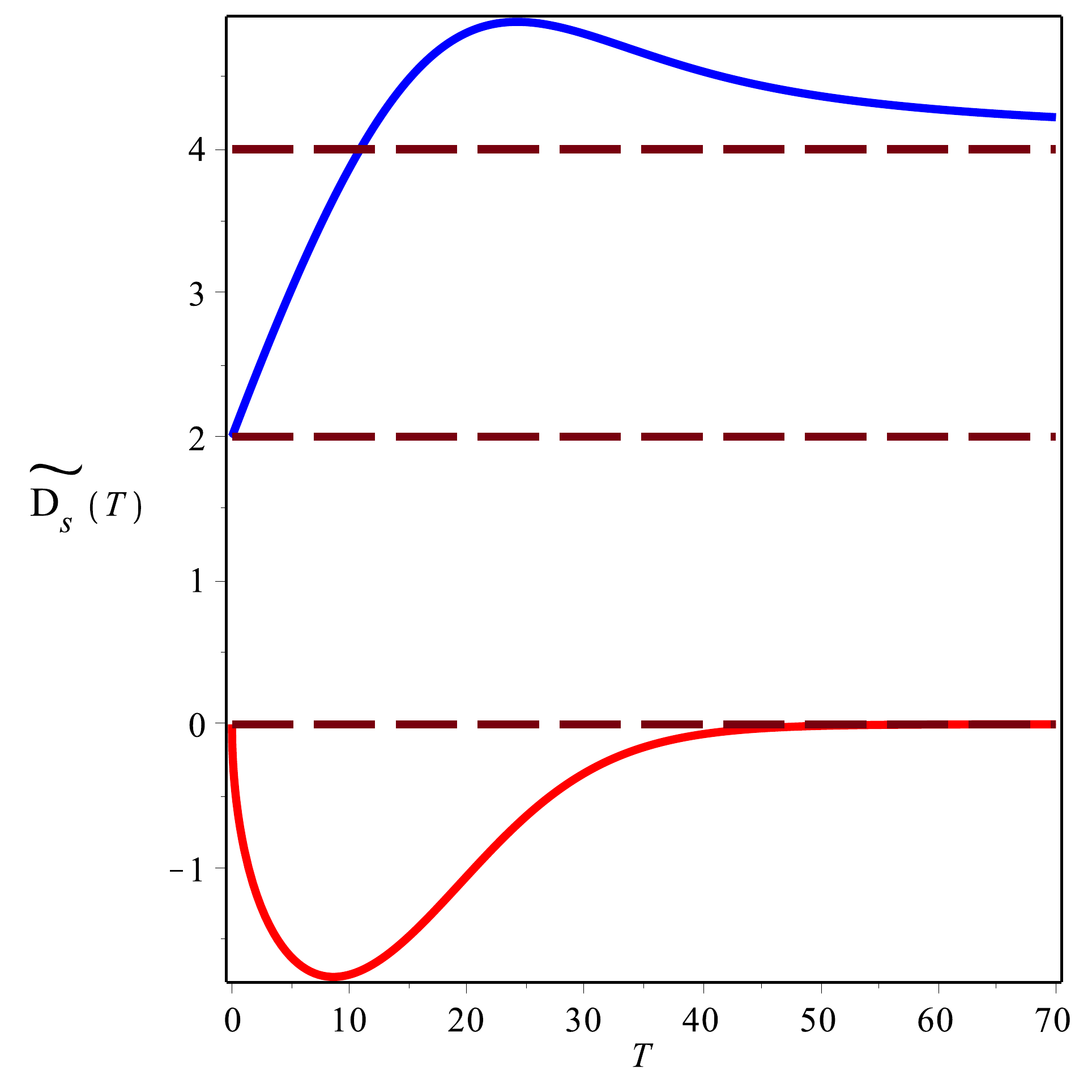}\end{center}
\caption{\sl Running spectral dimension $\widetilde {D_{\rm s}}(T)$ for our model. The parameter $a$ is chosen to be equal to one. The blue line represents the real part of $\widetilde {D_{\rm s}}(T)$,
while the red part represents the imaginary part. At $T\rightarrow 0$ the $\widetilde{D_{\rm s}}(T)$ approaches to the \emph{conventional} 
spectral dimension $D_{\rm s} = 2$, while at $T\rightarrow \infty$, it goes to the ``low energy" dimension $D_S^{\rm low} = 4$.
}
\end{figure}

Returning to the \emph{conventional} spectral dimension,
we see, that for all \emph{nonzero} real $a$ 
we get
\bea
D_{\rm s} \equiv\lim_{T\rightarrow 0} \widetilde {D_{\rm s}}(T) = 2~. \label{res}
\eea
Finally, although in the intermediate range of the parameter $T$
the spectral ``running" dimension is imaginary, there exists a sensible ``low energy " limit of $D_S$, 
valid again for all real $a$, with
\bea
D_{\rm s}^{\rm low} \equiv\lim_{T\rightarrow \infty} \widetilde {D_{\rm s}}(T) = 4~. \label{lowres}
\eea
Our result in Eq.~\eqref{res} is quite natural and the fact that the actual spectral dimension is 2, implies that it is in agreement
with the fact, that the gravitational propagators in our theory decrease faster at infinity due to the presence of the fourth 
derivative, thereby improving the ultraviolet  convergence of the Feynman loop diagrams. From another point of view our ``low energy''
result is in agreement with the fact that at very low energies the dynamics does not feel the Weyl square terms.

\noindent{{\bf Remark}\ \sl \small In principle relaxing the normalization condition discussed in the previous section, 
one can also renormalize the coefficient in front of the Weyl square action to a positive constant, that would correspond 
to negative $a$ in Eq.~\eqref{Sg2}. 
Such a renormalization would decrease a little bit the predictive power for curved
spacetime, however all particle phenomenology related with the flat spacetime would remain unchanged.
This normalization may be favorable in order to have positively defined expression (at large momenta) for
the Euclidean path integral, therefore for completeness we present also results for running dimension in this case.

In this situation, the running spectral dimension $\widetilde {D_{\rm s}}(T)$ is real for all $T$, but not just
at $T =0$ and $T = \infty$, and the corresponding plot is presented in Fig~\ref{figure3}. $\blacksquare$
\begin{figure}[top]\begin{center}
\includegraphics[scale = 0.4]{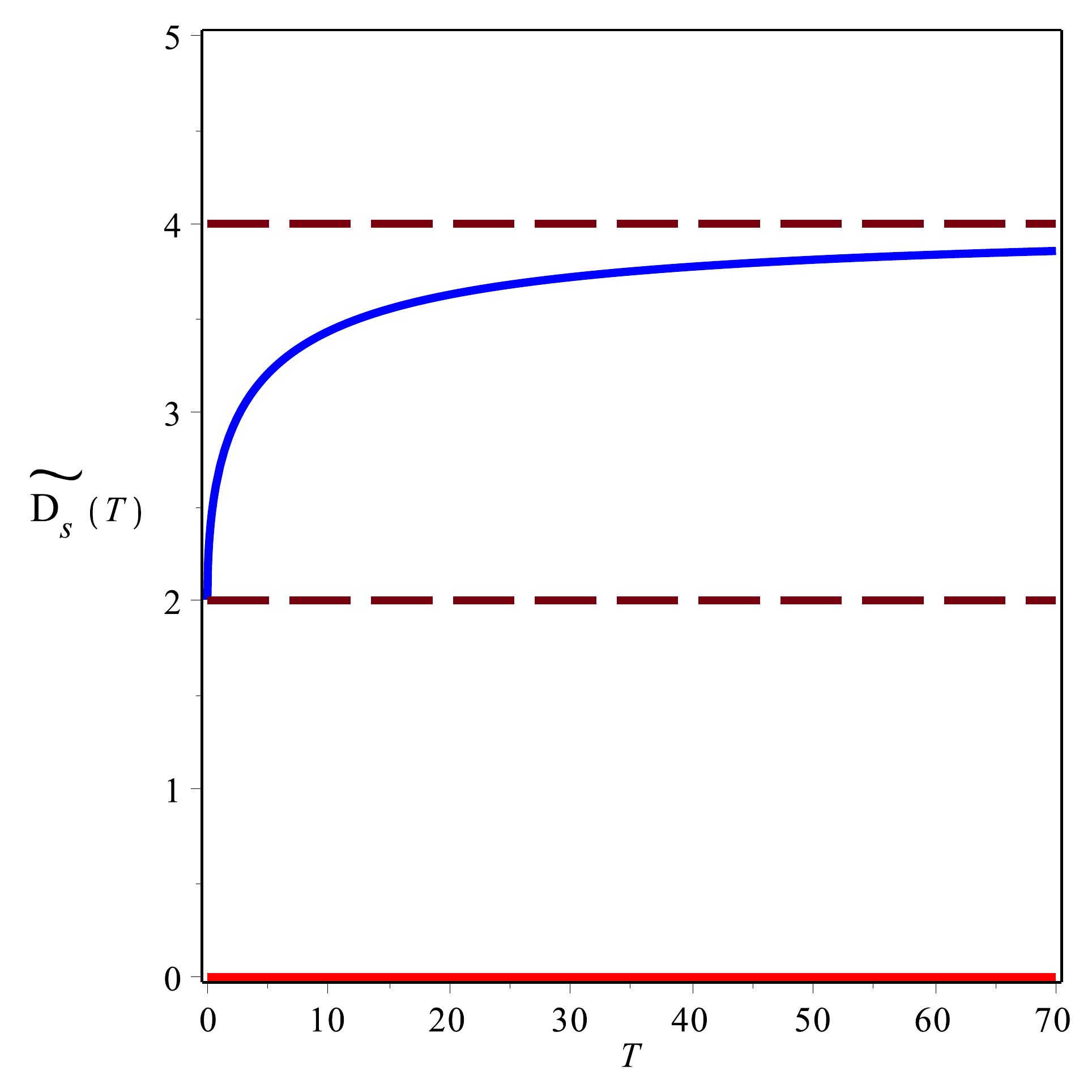}
\end{center}
\caption{\sl ``Running'' spectral dimension $\widetilde {D_{\rm s}}(T)$ for negative $a$ for our model. The parameter $a$ is chosen to be minus one. The  ``running" spectral dimension remains real for all $T$ from zero to plus infinity and represents at zero and at infinity
the same limits like for positive $a$.
}\label{figure3}
\end{figure}
}

\section{Conclusions and Outlook}
\label{last}
\setcounter{equation}{0}

In this paper we propose  a new definition of the bosonic spectral action using the zeta function regularization.  The corresponding theory is local, unitary and renormalizable. We recall that the mentioned statement concerns fermions and bosons, while quantization of gravity goes beyond the scope of our present study. Such a separation is indeed reasonable; quantum gravitational effects manifest themselves at energies $\sim M_{Pl} = 10^{19}$ GeV, but nonlocality and nonrenormalizability manifest theirselves at the cutoff scale $10^{14}-10^{17}$ GeV, i.e. at energies
\emph{at least} 100 times less. The spectral dimensions for fields of various spin are non trivial.

In order to obtain (in the ultraviolet renormalization) the Higgs quadratic term, a term in the Dirac operator corresponding to a neutrino Majorana mass is fundamental. A nonzero element in that position in the Dirac operator is also necessary to obtain the correct mass of the Higgs~\cite{Resilience}.  In this case the entry is a field
\be
a_i\psi^c \sigma(x) \psi, \quad i=1,2,3~, \la{mast}
\ee
(where $i$ is a generation index) and one  can consider more general terms 
\be
\psi^c (a_i\sigma(x) + M_i)\psi~, 
\ee
where $a_i$ and $M_i$ are different constants for right-handed neutrinos in different generations, with the condition $a_i/a_j\neq M_i/M_j$, for $i\neq j$, otherwise a field redefinition could eliminate them. These terms are allowed by symmetries\footnote{For
sterile fermions the mass terms  can be written either with constants or with a scalar field, hence also linear combinations are possible. The situation is qualitatively different from  left isospin doublets, where masses \emph{must } be generated via a scalar field.}. Indeed, the only reason of the usage of constant mass 
terms was the introduction of $M^4$, $M^2H^2$ and $M^2 R$ terms in the action and therefore the corresponding counter terms upon ultraviolet renormalization. The numerical values of the constants $M_i$ is not relevant, they can be arbitrarily small (and therefore without phenomenological consequences), but they must be nonzero in order to introduce the counterterms needed to renormalize the cosmological constant, the quadratic Higgs  and the Einstein-Hilbert terms. We emphasize, that at the present stage
all phenomenological predictions of our approach (like the prediction of the Higgs mass) are
the same as in the ones derived from the cutoff spectral action.

Working in the formalism exhibiting both $\sigma$ field and the first order condition, e.g.\ the grand symmetry framework~\cite{coldplay} , then 
the Dirac operator only has neutrino Majorana mass terms of the kind of \eqref{mast} and correspondingly no dimension zero and two operators appear in the classical action.
In this case the $\zeta$ spectral action reads:
\bea 
 &&  S_\zeta = \int dx \, \sqrt{g} \left( \gamma_1 B_{\mu\nu}B^{\mu\nu} + \gamma_2 W_{\mu\nu}^{\alpha}W^{\mu\nu\,\alpha} + \gamma_3 G_{\mu\nu}^aG^{\mu\nu\,a} 
+ \gamma_4\,H\left(-\nabla^2-\frac{R}{6}\right)H  \right. \nn
\\&& \left.+ \gamma_5 H^4+\gamma_6\,\sigma\left(-\nabla^2-\frac{R}{6}\right)\sigma + \gamma_7 \sigma^4+ \gamma_8 H^2\sigma^2 + \gamma_9 C_{\mu\nu\rho\sigma}C^{\mu\nu\rho\sigma} + \gamma_{10}R^*R^*\right), \label{SC} 
\eea
Both fermionic and bosonic parts of the spectral action are invariant under local conformal transformations
\be
\psi(x) \rightarrow e^{-\frac{3}{2}\phi(x)} \psi(x),\quad H(x) \rightarrow e^{-\phi(x)} H(x),\quad \sigma(x) \rightarrow e^{-\phi(x)} \sigma(x), \quad
e^{a}_{\mu}\rightarrow e^{+\phi(x)}e^{a}_{\mu},
\ee
and this classical theory does not contain any dimensionful parameters.  Since the theory is renormalizable 
these parameters will not appear in the renormalization process. 

A natural development of the theory described so far is the possibility to generate dynamically the three scales discussed above, thereby predicting them based on the spectral data and the unification point. Dynamical generation of scales upon quantization  has a long history dating back from Sakharov~\cite{Sakharov} for the gravitational sector, and Coleman and Weinberg~\cite{ColemanWeinberg} for the Higgs sector. The zeta spectral action~\eqref{SC} is a particular scale invariant extension of the scalar model. Such extensions are promising for the solution of the naturalness (hierarchy) problem~\cite{Scaleinvariantextension}. In this approach spontaneous symmetry breaking value is ``triggered'' by quantum correction and the Higgs vacuum expectation value can be computed~\cite{GildenerWeinberg}, thus increasing predictive power.
As far the gravitational sector is concerned, there are several examples where the gravitational constant can be induced, see for example the review~\cite{Visser}.  

It would be fascinating  to put a full fledged quantization of gravity in this scheme, nd even if this is still far in the future, we note that possible stating points could be conformal gravity, especially in the Bender-Mannhiem formalism \cite{ Mannheim:2006rd,Bender:2007wu,Bender:2008gh}, which leads to unitary and renormalizable theory of gravity \cite{Mannheim:2011ds} and has interesting astrophysical consequences \cite{Mannheim:2010ti,Mannheim:2010xw}.
 Interesting connections between conformal and Einstein gravities are discussed in Ref.~\cite{Maldacena}. 

In conclusion, the zeta spectral action is an interesting alternative  to the usual  cutoff spectral action. It shows promises of explaining the phenomenology of the Standard Model and beyond. In addition, the way it treats the fundamental scales could also shed some light on the explanation of some fundamental questions.

\noindent{\bf Acknowledgements}  We thank O.O.~Novikov and P.~Mannheim for valuable discussions and correspondence,
and also  F. D'Andrea and W. van Suijlekom for useful comments. F.L. is partially supported by CUR Generalitat de Catalunya under projects FPA2013-46570 and 2014~SGR~104. M.A.K.\ and F.L.\ were partially supported by UniNA and Compagnia di San Paolo under the grant Programma STAR 2013. 
M.A.K. is partially supported by RFBR grant 13-02-00127-a. A.W. is partially supported by a Royal Thai Government Scholarship.

\end{document}